\documentclass[prb,twocolumn,showpacs]{revtex4}
\usepackage{srctex}
\usepackage{amssymb}
\usepackage{amsmath}
\usepackage{epsfig}
\usepackage{graphicx}
\usepackage{dcolumn}
\usepackage{bm}
\usepackage[dvipsnames]{xcolor}

\begin{document}

\date{\today}
\title{\textbf{Boson gas in a periodic array of tubes}}
\author{P. Salas$^{1,2}$, F.J. Sevilla$^2$ and M.A. Sol\'{\i}s%
$^2$}
\address{$^1$Posgrado en Ciencia e Ingenier\'{\i}a de Materiales, UNAM, \\
Apdo. Postal 70-360, 04510 M\'exico, D.F., MEXICO\\
$^2$Instituto de F\'{\i}sica, UNAM, Apdo. Postal 20-364,
01000 M\'exico, D.F., MEXICO
}

\begin{abstract}
We report the thermodynamic properties of an ideal boson gas confined in an infinite periodic array of channels modeled by two, mutually perpendicular,
Kronig-Penney delta-potentials. The particle's motion is hindered in the $x$-%
$y$ directions, allowing tunneling of particles through the walls, while no confinement along the $z$ direction is considered.
It is shown that there exists a finite Bose-Einstein condensation (BEC)
critical temperature $T_{c}$ that decreases monotonically from the 3D ideal
boson gas (IBG) value $T_{0}$ as the strength of confinement $P_{0}$ is increased
while keeping the channel's cross section, $a_x  a_y$ constant. In contrast, $%
T_{c}$ is a non-monotonic function of the cross-section area for fixed $P_{0}
$. In addition to the BEC cusp, the specific heat exhibits a set of maxima
and minima. The minimum located at the highest 
temperature is a clear signal of the confinement effect which occurs when the boson wavelength is twice the cross-section side size. This confinement is amplified when the wall strength is increased until a dimensional crossover from 3D to 1D is produced. 
Some of these features in the specific heat obtained from this simple model can be
related, qualitatively, to at least two different experimental situations: $%
^{4}$He adsorbed within the interstitial channels of a bundle of carbon 
nanotubes and superconductor-multistrand-wires Nb$_{3}$Sn.
\end{abstract}

\maketitle

\section{Introduction}

The properties of quantum systems at low dimensionality and temperature have
attracted the attention of researchers for a long time. In particular, the
study of phase transitions, such as superfluidity, superconductivity, or
Bose-Einstein condensation, has been marked by an impressive interest among
scientists in the field even though ``true'' long-range order phases are excluded
by the Mermin-Wagner-Hohenberg theorem \cite{MerminPRL66,HohenbergPR67} in
dimensions lower or equal than two.

In two dimensions, however, the superfluid or superconductor transition arises
by the acquisition of quasi-long-range-order in the system as described by
the Kosterlitz-Thouless phase transition \cite%
{KosterlitzJPC73,KosterlitzJPC74}. This, has been experimentally confirmed
in thin superconducting- \cite%
{HebardPRL80,BenfattoPRL2007,HalperinJLTP79,Turkevich,DoniachPRL79,String94}
and helium-films \cite{BishopPRL78,Agnolet,LuhmanPRL2004}.
Historically, different experimental methods have been set up to study low
dimensional phase transitions since the discovery of helium superfluidity.
Indeed, studies on the behavior of helium films have been performed by using
several experimental techniques, like the adsorption of helium on simple
plane substrates \cite{Lauter-Hefilm}, or even on more complex ones as in
nanoporous media such as cylindrical pores of Anopore \cite{Finotello},
Vycor \cite{Vycor} or Gelsil \cite{Gelsil} glasses, where the effects of the
substrate structure on the specific heat has been reported among other
properties. More recently, studies of adsorption of atoms and molecules on
planar substrates \cite{DashBook,BrunchBook}; the two-dimensional character
of high critical temperature superconductivity; and the discovery of
graphene, have made two-dimensional systems to be widely explored contrary
to the case of quasi-one-dimensional ones.

Although the theoretical aspects of one-dimensional systems have been
extensively studied \cite{Lieb}, only very recently 1D experimental reports
have attracted a great attention. For example, now days it is possible to create a 1D Bose gas in cigar-shaped
magneto-optic traps \cite{cigar-traps} where the particle density, the cigar
size and the intensity of the interaction between particles are
experimentally tunable parameters, allowing one to examine quantum phenomena
such as
the superfluid to Mott-insulator phase transitions \cite{Greiner1D,1D-Mott-superfluid}.

On the other hand, 
with the advent of carbon nanotubes, the realization of phase transitions in
quasi-one-dimensional systems of different substances adsorbed on nanotube
bundles is now possible \cite{Iijima,Thess,Teizer,WangSciende2010}. The quasi-one-dimensional
character of these structures is a consequence of the enormous aspect ratios
that nanotubes exhibit, with cross-sections in the nanoscale regime. On
such length scales the single-particle energy levels corresponding to the
cross-section degrees of freedom are \textquotedblleft
frozen\textquotedblright\ leading to effective one-dimensional systems. 

In some other recent theoretical studies \cite{AncilottoPRB2004,MarconePRB2006}, the occurrence of BEC of a
weakly-interacting quantum gas of Bose particles (parahydrogen or $^{4}$He)
adsorbed within the interstitial channels (IC) of a bundle of poly-disperse
carbon nanotubes has been predicted. The reported BEC transition and
particularly, the dependence of the specific heat on temperature, exhibit
features of four dimensions in contrast to the expected one-dimensional
behavior that has been, indeed, observed in the experimental report of
Lasjaunias \textit{et al.} \cite{Lasjaunias2003} of the specific heat of
adsorbed $^{4}$He in nanotubes. The authors of the former references, Refs. [%
\onlinecite{AncilottoPRB2004,MarconePRB2006}], argue that the presence of
\textit{nonuniformity} in the nanotube cross-section gives rise to three
additional degrees of freedom (the radii of the three tubes that form the IC),
needed in their analysis to ensure the occurrence of 
the BEC transition which, 
as they claim \cite{Gatica} based on the results reported in Ref. [\onlinecite{ShiPRL2003}],
doesn't exist in a uniform bundle of identical one-dimensional nanotubes.

Truly, it is well known that there is no BEC of a non-interacting boson gas in an
impenetrable one-dimensional box potential. Also, that there is no BEC whenever
the spatial dimensions of at least one direction is finite, thus excluding
the possibility of BEC in just one channel. Therefore, a collection of independent IC's can not develop a BEC unless a coupling mechanism, such as tunnel-effect, is present between adjacent channels. A possible mechanism that couple the different IC's is considered in Refs. [\onlinecite{AncilottoPRB2004,MarconePRB2006}], where the authors argue that such a coupling leads to an effective density of states from which a non-vanishing BEC critical temperature is obtained. This effective density of states  can be described as an inhomogeneously broadened convolution of the density of the heterogeneous transverse states with the one-dimensional density of states of a free particle that moves along the nanotube axis.  

Here we suggest that the packing defects due to the heterogeneity of the nanotube cross-section \cite{ShiPRL2003} may lead to a tunnel-effect-like mechanism that couples the different interstitial channels in the bundle, thus making BEC possible. 
Indeed, it is well known that suitable confinement potentials can make the BEC long-range-order character to be
stable against long-range-fluctuations. Such is the case of the ideal Bose
gas in two dimensions trapped by harmonic potentials, which undoubtedly
undergoes BEC. So, even in the case of a collection of nanotubes with uniform
cross-section, one would expect a BEC transition at a finite critical
temperature when tunneling between neighboring channels (modeled using penetrable potential barriers) is considered.

In addition to the theoretical interest, the model presented in this paper
can be applied to estimate critical temperatures and thermodynamic
properties of superconductor-multistrand-wires whose technological
applications, that go from Nuclear Magnetic Resonances to magnets used in
high-energy accelerators, impel their understanding.
For example, several authors report experimental studies on the
thermodynamic properties of multistrand Nb$_{3}$Sn (an A15 type of
superconductor) wires, that exhibit a transition temperature around 18 K and
can support fields up to 15 Tesla  \cite{Junod2007}. Typical numbers of filaments range from $%
10^{2}$ to $10^{4}$ Nb$_{3}$Sn superconductor wires with diameters varying
from a few to tens $\mu $m.
On the other hand, Nb$_{3}$Sn multifilament wires come in different sizes,
shapes and compositions, depending on the techniques used to create them.
The most common are the Bronze Route, the Internal Sn diffusion process and
the Powder Metallurgy (PM) methods, and they may have a core either of Cu, Sn, NbSn or NbCu alloys, while usually immersed in Cu, and the use of tantalum and/or titanium barriers to prevent the Sn pollution in Cu  \cite{Miyazaki}.
Experimentalists report either
total specific heat curves and/or curves that subtract the normal state
specific heat, to avoid phonon and unpaired electron interference \cite%
{Junod2007,Wang2006,WAMSDO}. The main feature in this curves is the
transition around 18 K with a typical width of 5 K. However, in some cases a
second peak appears around 9 K which the authors interpret as the trasition of the remnants of
unreacted Nb. Here, in light of our results we suggest an alternative interpretation to the meaning of these maxima.  

In this paper we report the thermodynamic properties with emphasis in the BEC critical temperature and the specific heat, of an ideal 
Bose gas within an infinite periodic array of tubes. Our results are benchmarks for ongoing studies on the properties of real spatially confined systems such as: a) He atoms in interstitial carbon nanotube bundles, b)
Cooper pairs in periodic tubes like multistrand Nb$_{3}$Sn bundles or
Bechgaard-salts \cite{Bechgaard} or c) bosonic atoms in two dimensional
opto-magnetic traps \cite{Greiner1D}.

Although at very low (or zero) temperatures and/or high densities the
interaction between particles cannot be neglected, we focus on an
interactionless boson gas to study the effects of a periodic confining
potential on the properties of the system. We show that this simple model
captures qualitatively the properties of real systems, including the
emergence of thermal phase transitions and/or dimensional crossovers \cite%
{PatyJLTP,PatyPRA}.

In the following section we describe our system model. In Sec. III we
calculate the Bose-Einstein condensation critical temperature in addition to
the specific heat and other relevant thermodynamic variables. In Sec. IV we discuss the results and 
present our conclusions.

%
%
%

\section{Periodic tube bundles}

\begin{figure}[tbh]
\centerline{\epsfig{file=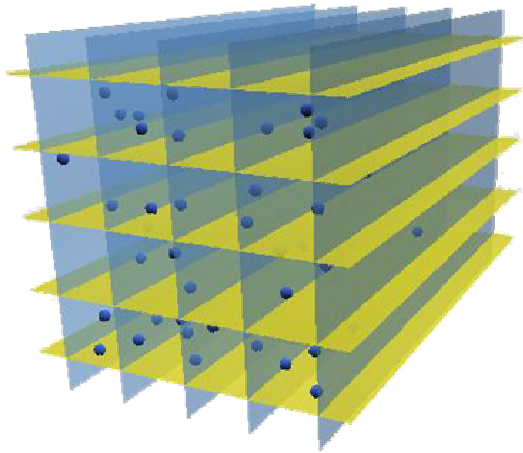,height=2.8in,width=2.4in,angle=90}}
\caption{(Color online) Periodic array of square cross section tubes.}
\label{fig:Tubos}
\end{figure}

Our system model consists of $N$ non-interacting bosons confined in an infinite
periodical array of penetrable tubes of rectangular cross section of sides $%
a_{x}$ and $a_{y},$ and infinite length. We model the tubes array by
considering two perpendicular Kronig-Penney (KP) delta barriers in the $x$
and $y$ directions with no constraints in the remaining $z$ direction (see
Fig. \ref{fig:Tubos}). Our periodic structure resembles either the bundle of
homogeneous nanotubes, the superconductor-multistrand-wires or the experimental 2D
periodic lattice of tightly confined potential tubes created in Ref. [\onlinecite{Greiner1D}]. 
If the interaction between bosons is ignored, the
Schr\"{o}dinger equation for each boson of mass $m$ in this system is%
\begin{equation}
\left\{ -\frac{\hbar ^{2}}{2m}\nabla ^{2}+ V(x,y) \right\} \psi (x,y,z) =
\varepsilon _{k}\psi (x,y,z)  \label{Schre}
\end{equation}
with
\begin{equation}
V(x,y) = \sum_{n=-\infty }^{\infty }v_{x}\delta (x-na_{x})+\sum_{n=-\infty
}^{\infty }v_{y}\delta (y-na_{y})  \label{Vxy}
\end{equation}%
where $v_{x}$ and $v_{y}$ are the delta strength in the $x$ and $y$
directions, respectively.

The Schr\"{o}dinger equation (\ref{Schre}) is separable in each direction
such that $\varepsilon _{k}=\varepsilon _{k_{x}}+\varepsilon
_{k_{y}}+\varepsilon _{k_{z}}$ is the energy per particle, where
\begin{equation}
\varepsilon _{k_{z}}=\frac{\hbar ^{2}k_{z}^{2}}{2m},  \label{Green0}
\end{equation}%
with $k_{z}=2\pi n_{z}/L$ the momentum in the $z$-direction, $n_{z}=0,\pm
1,\pm 2,...$ due to the periodic boundary conditions in a box of length $L$,
and $\varepsilon _{k_{x,y}}$ are implicitly obtained from the equations \cite{KP}
\begin{equation}
(P_{i}/\alpha _{i}a_{i})\sin (\alpha _{i}a_{i})+\cos (\alpha _{i}a_{i})=\cos
(k_{i}a_{i}),  \label{KPRD}
\end{equation}%
with $\alpha _{i}^{2}\equiv 2m\varepsilon _{k_{i}}/\hbar ^{2}$, where $i=x$
or $y$. We rewrite the dimensionless constants $P_{i}=mv_{i}a_{i}/\hbar ^{2}$
as $P_{i}=(mv_{i}\lambda _{0}/\hbar ^{2})(a_{i}/\lambda _{0})\equiv
P_{0i}(a_{i}/\lambda _{0})$, where $\lambda _{0}\equiv h/\sqrt{2\pi
mk_{B}T_{0}}$ is the de Broglie thermal wavelength of an ideal boson gas in
an infinite box at the critical temperature $T_{0}=2\pi \hbar
^{2}n_{B}^{2/3}/mk_{B}\zeta (3/2)^{2/3}\simeq 3.31\hbar
^{2}n_{B}^{2/3}/mk_{B}$ with $n_{B}\equiv N/L^{3}$ the boson number density
and $a_{i}$ the distance between the deltas barriers along the $i=x$ and $y$
directions. $P_{0i}$ is a measure of the tube wall impenetrability directly
related to the delta-barrier strength. This can be understood if we recall
the transmission coefficient for an one dimensional delta potential of
strength $v_{i}$, $\tau _{i}=1/(1+P_{0i}^{2}/\bar{E}),$ \cite{deLlano} with $%
P_{0i}$ as defined above and $\bar{E}$ is the energy, in $\hbar ^{2}/2ma^{2}$
units, of particles arriving at right angles with the delta potential. We
recover the following two limits: when $P_{0i}$ goes to infinity the
transmission coefficient vanishes and our model becomes an infinite number
of decoupled tubes; when $P_{0i}=0$, $\tau _{i}=1$ and the confining tube
walls disappear recovering the 3D ideal Bose gas. By performing a series
expansion of the left-hand-side of (\ref{KPRD}) just above of the exact $%
P_{i}$-dependent single-particle-ground-state energy $\varepsilon _{0i}$ in
the first band, the single-particle energy spectrum can be written to first
order as
\begin{equation}
\varepsilon _{k_{i}}\simeq \varepsilon _{0i}+\frac{\hbar ^{2}}{M_{i}a_{i}^{2}%
}(1-\cos k_{i}a).  \label{eqWK}
\end{equation}%
This is the dispersion that has been considered by several authors \cite%
{wenkan,haerdig} in the standard nearest-neighbor hoping approximation of
the well-known Hubbard model with $\varepsilon _{0i}=0.$
%
The effective mass $M_{i}$ is given by the expression $m[(\alpha
_{0i}a_{i})^{-1}\sin \alpha _{0i}a_{i}+P_{i}(\sin \alpha _{0i}a_{i}-\cos
\alpha _{0i}a_{i})(\alpha _{0i}a_{i})^{-3}]$ with $\alpha _{0i}a_{i}=\sqrt{%
2m\varepsilon _{0i}/\hbar ^{2}}$ and valid for $P_{i}\gtrsim 0.06$. As
expected, $M_{i}$ grows when $P_{i}$ increases and for $P_{i}\gtrsim 40$ the
relation is almost linear. 
Even though (\ref{eqWK}) might seem to be a good approximation to calculate the thermodynamic properties of the system \cite{haerdig} in the low temperature regime, there are distinct effects, 
particularly in the specific heat, that can only be observed when the full
band spectrum is considered as is shown below. 

\section{Critical temperature and Specific heat}

\subsection{Grand potential}

The thermodynamic properties of the system are obtained from the grand
potential $\Omega (T,L^{3},\mu )$ for a boson gas
\begin{gather}
\Omega (T,L^{3},\mu )=U-TS-\mu N  \notag \\
=\Omega _{0}+k_{B}T \sum_{\mathbf{k} \neq 0}\ln [1-e^{-\beta (\varepsilon
_{k_{i}}-\mu )}]  \label{omega}
\end{gather}%
where $\Omega _{0}$ is the contribution of the ground state $k_{i}=0$ with $%
i=x,\,y,\,z$, and $\beta \equiv 1/k_{B}T$. As usual, $U,$ $S$ and $\mu$ denote the internal energy, the entropy and the chemical potential respectively. By using the dispersion relations
given by (\ref{Green0}) and (\ref{KPRD}) and after some algebra we obtain
\begin{eqnarray}
&&\hspace{-0.4cm}\Omega \left( T,L^{3},\mu \right) =k_{B}T\ln [1-e^{-\beta
(\varepsilon _{0}-\mu )}]-\frac{L^{3}m^{1/2}}{\left( 2\pi \right)
^{5/2}\hbar }\frac{1}{\beta ^{3/2}}\times   \notag \\
&&\int_{-\infty }^{\infty }\int_{-\infty }^{\infty }dk_{x}\
dk_{y}g_{3/2}(e^{-\beta (\varepsilon _{k_{x}}+\varepsilon _{k_{y}}-\mu )}),
\label{TGP}
\end{eqnarray}%
where we have replaced the summations by integrals ${\ \sum_{\mathbf{k}%
}\longrightarrow }(L/2\pi )^{3}\int d^{3}\mathbf{k}$, assuming 
$\hbar ^{2}/mL^{2}\ll k_{B}T$ , and we have introduced the
Bose functions \cite{Path} $g_{\sigma }(t)\equiv \sum_{l=1}^{\infty }{t{^{%
\mathit{l}}}/{\mathit{l}^{\sigma }}}$. $\varepsilon _{0}=\varepsilon _{0x}+ \varepsilon _{0y}$ is the ground
state energy which depends on $P_{0i}$ and on $a_{i}/\lambda _{0}$.

From (\ref{TGP}) the thermodynamic properties for a monoatomic gas can be
calculated using the relations
\begin{eqnarray}
N=-\left( {\frac{\partial \Omega }{\partial \mu }}\right) _{T,L^{3}}, && \hspace{%
-0.4cm}U(T,L^{3})=-k_{B}T^{2}\left[ {\frac{\partial }{\partial T}}\left( {\frac{%
\Omega }{k_{B}T}}\right) \right] _{L^{3}, \mathsf{z}}  \notag \\
\mbox{and}\ \ C_{V} &=&\left[ {\ \frac{\partial }{\partial T}}U(T,L^{3})\right]
_{N,L^{3}}.  \label{edos2}
\end{eqnarray}%
where $\mathsf{z} \equiv \exp (\beta \mu) $ is the fugacity.

\subsection{Critical temperature}

We define the critical temperature $T_{c}$ as the temperature when the
number of bosons in the ground-state level ceases to be negligible, i.e.,
$N_{0}(T_{c})\simeq 0$ and the chemical potential $\mu(T_{c}) \simeq \mu _{0}%
=\mu _{0x}+\mu _{0y} = \varepsilon_0$. 

From the first expression in Eq. (\ref{edos2}) and Eq. (\ref{TGP}) we obtain
the particle number $N$
\begin{eqnarray}
N &=&\frac{1}{e^{\beta (\varepsilon _{0}-\mu )}-1}+L^{3}\sqrt{\frac{m}{2\pi
^{5}\hbar ^{2}\beta}}\times  \notag \\
&&\int_{0}^{\infty }\int_{0}^{\infty }dk_{x}\ dk_{y}g_{1/2}(e^{-\beta
(\varepsilon _{k_{x}}+\varepsilon _{k_{y}}-\mu )}).  \label{num}
\end{eqnarray}

At $T=T_{c}$, the first term vanishes and the critical temperature is
obtained from
\begin{eqnarray}
\frac{N}{L^{3}} &=&\sqrt{\frac{m}{2\pi ^{5}\hbar ^{2}\beta _{c}}}%
\int_{0}^{\infty }\int_{0}^{\infty }dk_{x}\ dk_{y}\times  \notag \\
&&g_{1/2}(e^{-\beta _{c}(\varepsilon _{k_{x}}+\varepsilon _{k_{y}}-\mu
_{0})}).
\end{eqnarray}
Here we set the number density $N/L^3$ equal to that of an ideal Bose gas (IBG) in the
thermodynamic limit, with a BEC critical temperature $T_0$.
Note that all the integrals involving the energy-spectrum in the $x$ and $y$ directions can be split in a sum of integrals over the energy bands folded in the first Brillouin zone. 

\begin{figure}[tbh]
\epsfig{file=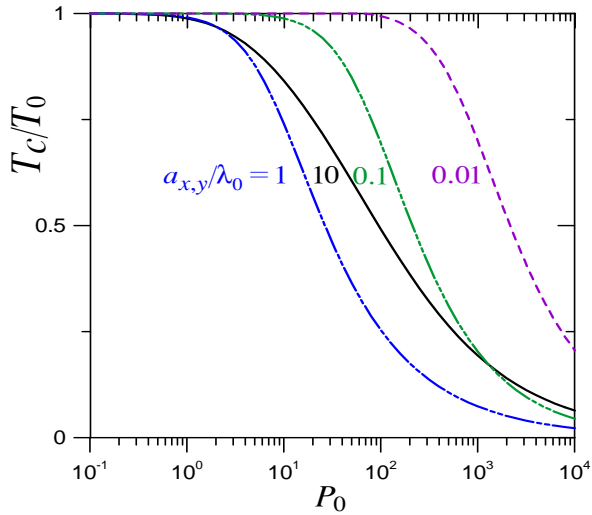,height=2.7in,width=3.2in}
\caption{(Color online) Critical temperature in units of $T_0$ as a function
of $P_0$ for different values of $a_{x}/\protect\lambda _{0}= a_{y}/\protect%
\lambda _{0}=a/\protect\lambda _{0}$.}
\label{fig:Tc1}
\end{figure}

\begin{figure}[tbh]
\centerline{\epsfig{file=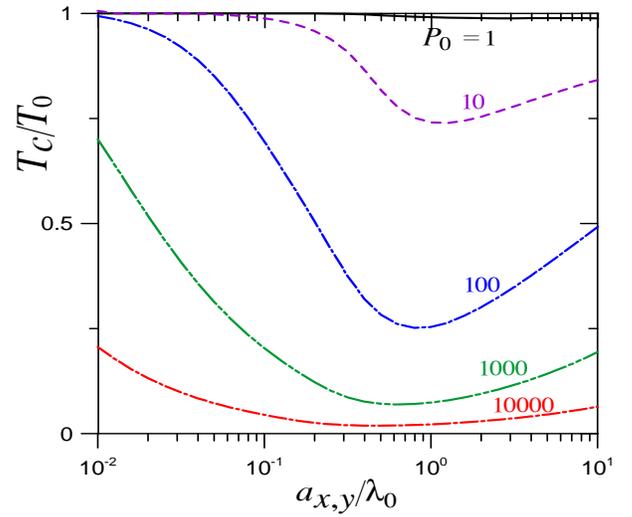,height=2.7in,width=3.3in}}
\caption{(Color online) Critical temperature in units of $T_0$ as a function
of $a_{x}/\protect\lambda _{0}=a_{y}/\protect\lambda _{0}=a/\protect\lambda %
_{0}$, for different values of \ $P_{0}$.}
\label{fig:Tc2}
\end{figure}

In the {\it isotropic} case, where $P_{0x}=P_{0y}\equiv P_0$ and $a_x=a_y\equiv a$,
the critical temperature as a function of the parameter $P_{0}$ is shown in
Fig. \ref{fig:Tc1} for different values of the tube cross-section. Note that as
the impermeability $P_{0}$ of the tube walls increases, the critical temperature
diminishes monotonically from $T_0$. In contrast, the variation of $T_{c}$ as function of $a/\lambda _{0}$ shows a non-monotonic behavior (Fig. \ref{fig:Tc2}), decreasing for $a/\lambda_{0}\lesssim1$ and increasing otherwise .


For both Figs. \ref{fig:Tc1} and \ref{fig:Tc2}, we find a similar qualitative
behavior of the critical temperature as that reported for a boson gas in multilayers \cite{PatyJLTP,PatyPRA},
namely a decrease in $T_c/T_0$ as $P_0$ increases and a trend of $T_c/T_0$
to go back to unity as $a/\lambda _{0}$ increases after having reached a
minimum. However, we notice that for similar $P_0$ and $a$ values as those
used in multilayers, we obtain even lower $T_c/T_0$ values for bosons in square cross-section tube bundles , 
showing that the presence of an additional KP delta potential in the
two-dimensional array emphasizes even more the effects of confinement.
Although we have only shown the critical temperatures for the isotropic case, we will show in the following sections, the effects of anisotropy in the internal energy and specific heat by considering a rectangular cross-section.

\subsection{Internal Energy}

The temperature-dependent internal energy $U$ is given by
\begin{eqnarray}
U-\varepsilon _{0}N &=&L^{3}\sqrt{\frac{m}{2\pi ^{5}\hbar ^{2}\beta}}
\int_{0}^{\infty }\int_{0}^{\infty }dk_{x}\ dk_{y} \times  \notag \\
&& \left\{ (\varepsilon -\varepsilon _{0})g_{1/2}(\chi ) +\frac{1}{2\beta }%
g_{3/2}(\chi )\right\} ,  \label{IUU}
\end{eqnarray}%
where $\varepsilon \equiv \varepsilon _{k_{x}}+\varepsilon _{k_{y}}$ and $%
\chi \equiv e^{-\beta (\varepsilon -\mu )}.$

\begin{figure}[tbh]
\centerline{\epsfig{file=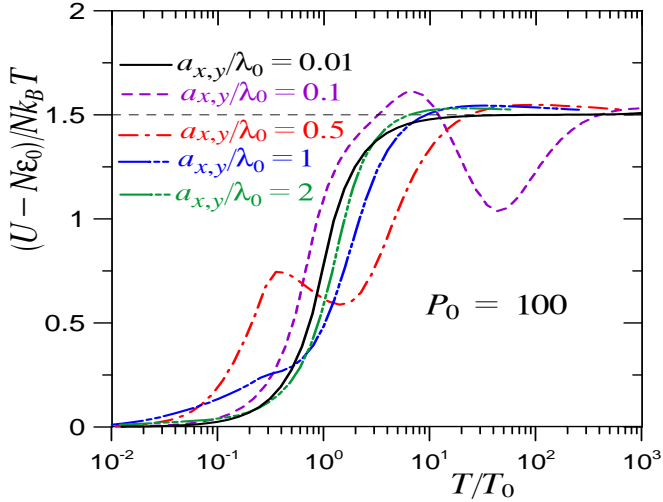,height=2.7in,width=3.5in}}
\caption{(Color online) Isotropic case. Internal energy as a function of $%
T/T_0$, for tube arrays of square cross section of several side sizes $a_x/%
\protect\lambda_0=a_y/\protect\lambda_0$ and $P_0=100$.}
\label{fig:U1}
\end{figure}

\begin{figure}[tbh]
\centerline{\epsfig{file=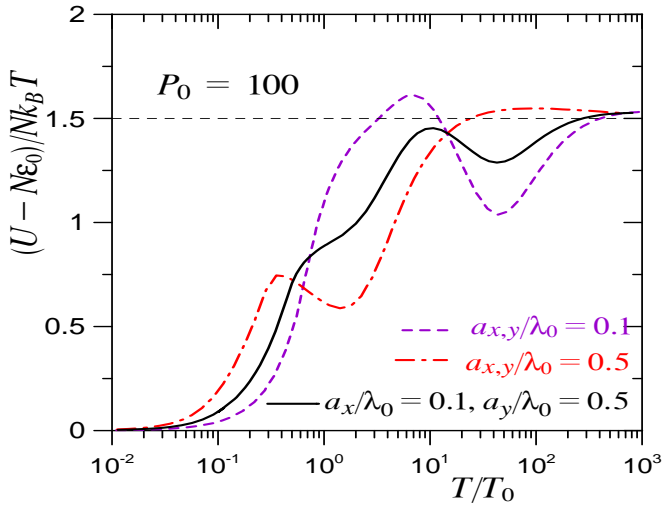,height=2.7in,width=3.5in}}
\caption{(Color online) Anisotropic case. Internal energy as a function of $%
T/T_{0}$, $a_{x}/\protect\lambda _{0}=0.5$, $a_{y}/\protect\lambda _{0}=0.1$
and $P_{0}=100$ (black line.) For comparison, the isotropic curves are also
shown (color lines).}
\label{fig:U2}
\end{figure}

In Fig. \ref{fig:U1} we show the internal energy referred to the system
ground state energy, for a square cross section, $a_{x}=a_{y}$, and in Fig. %
\ref{fig:U2} we show results when $a_{x}$ is different from $a_{y}$. %
%
%
In the isotropic case, the internal energy curves increase monotonically for
$a_{x,y}$ larger than $\lambda_0$. However, when the separation is $%
a_{x,y}/\lambda _{0}\simeq 0.5$ or lower, the effect of the KP potentials is
revealed by two maxima that become more pronounced as the tube cross-section
decreases. This behavior of the internal energy is similar, albeit more
complex, in the anisotropic case where the maxima and minima of the
corresponding isotropic cases are still present.
Minima are associated with the loss of freedom of particles in two
directions, which we will discuss in more detail when we analyze the
specific heat behavior.

\subsection{Specific heat}

From Eq. (\ref{edos2}) and (\ref{IUU}), the specific heat becomes
\begin{eqnarray}
&& \hspace{-0.6cm} \frac{C_{V}}{Nk_{B}} =\frac{L^{3}}{N}\sqrt{\frac{m\beta }{%
8\pi ^{5}\hbar ^{2}}}\int_{0}^{\infty }\int_{0}^{\infty }dk_{x}\ dk_{y}\left[
g_{1/2}(\chi )\right.  \notag \\
&& \hspace{3.1cm}\times (2\varepsilon -\varepsilon _{0}-\mu +T\frac{d\mu }{dT%
})  \notag \\
&&\hspace{-1.0cm}\left. +2\beta (\varepsilon -\varepsilon _{0})g_{-1/2}(\chi
)(\varepsilon -\mu +T\frac{d\mu }{dT})+\frac{3}{2\beta }g_{3/2}(\chi )\right]
\label{cvt}
\end{eqnarray}%
For $T<T_{c}$ the chemical potential $\mu =\mu _{0}$ is a constant, $%
\partial \mu /\partial T=0$ and, using $\chi _{0}\equiv e^{-\beta
(\varepsilon -\mu _{0})}$ the last equation for the specific heat becomes
\begin{eqnarray}
&& \hspace{-0.50cm} \frac{C_{V}}{Nk_{B}} =\frac{L^{3}}{N}\sqrt{\frac{m\beta
}{2\pi ^{5}\hbar ^{2}}}\int_{0}^{\infty }\int_{0}^{\infty }dk_{x}\
dk_{y}\left\{g_{1/2}(\chi _{0})(\varepsilon -\mu _{0}) \right.  \notag \\
&& \left. +\beta (\varepsilon -\mu _{0})g_{-1/2}(\chi _{0})(\varepsilon -\mu
_{0}) +\frac{3}{4\beta }g_{3/2}(\chi _{0})\right\}.  \label{cvt00}
\end{eqnarray}%
%
%
%
%
\begin{figure}[tbh]
\centerline{\epsfig{file=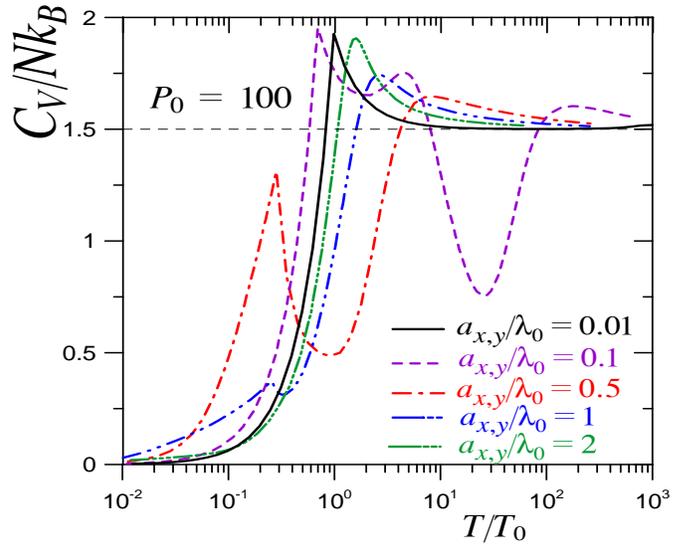,height=2.9in,width=3.5in}}
\caption{(Color online) Isotropic case. Specific heat in $Nk_{B}$ units, as
a function of $T/T_{0}$, for different $a_{x}=a_{y}$ values and $P_{0}=100$.}
\label{fig:Cv1}
\end{figure}
%
\begin{figure}[tbh]
\centerline{\epsfig{file=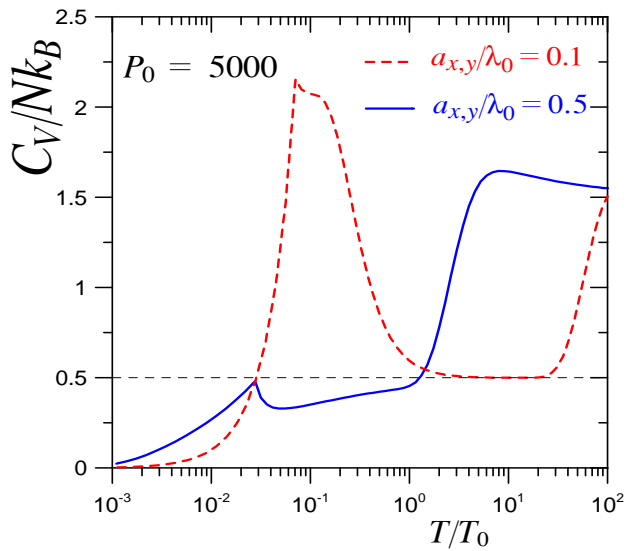,height=2.9in,width=3.3in}}
\caption{(Color online) Specific heat in $Nk_{B}$ units, as a function of $%
T/T_{0}$, for $P_{0}=5000$, $a_{x}/\protect\lambda _{0} = a_{y}/\protect%
\lambda _{0} = 0.5$ and $0.1$}
\label{fig:CvP05000}
\end{figure}

\begin{figure}[tbh]
\centerline{\epsfig{file=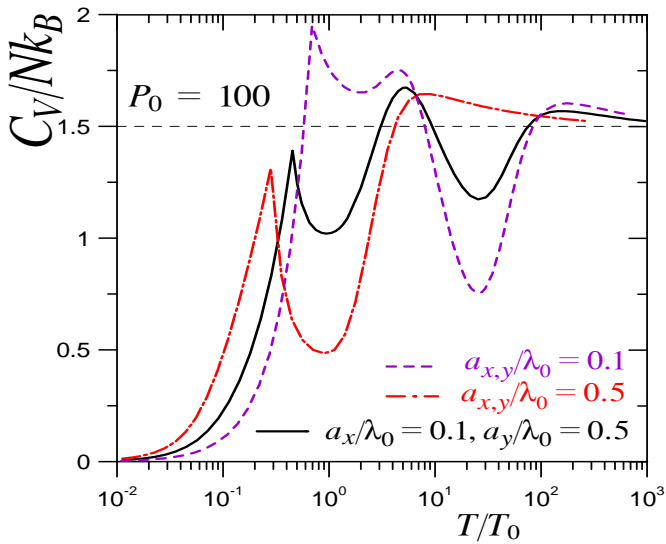,height=2.9in,width=3.5in}}
\caption{(Color online) Anisotropic case. Specific heat in $Nk_{B}$ units,
as a function of $T/T_{0}$, for $a_{x}/\protect\lambda _{0}=0.5$ and $a_{y}/%
\protect\lambda _{0}=0.1$ and $P_{0}=100$ (black line) compared to their respective
isotropic cases: $a_{x}/\protect\lambda _{0}=a_{y}/\protect\lambda _{0}=0.1$
and $0.5$, (color lines).}
\label{fig:Cv2}
\end{figure}
In Fig. {\ref{fig:Cv1}} we show the specific heat for the isotropic case, as a function of $T/T_{0}$ for different values of $a_{x}=a_{y}=a$ and $P_{0}$ fixed to $100$. The BEC critical temperature of the system 
is displayed as a sharp peak in the specific heat that occurs at temperature whose value is lower than the critical temperature of a free ideal Bose gas. Except for the case with 
$a_x/\lambda_{0}=0.1$ where two minima appear, the specific heat shows only one minimum at a characteristic temperature that reveals the only length scale of the system, namely $a_{x}=a_{y}.$ Such temperature satisfies the relation $\lambda\simeq2a_{x}$ and marks the maximum effects of confinement, where $\lambda =\left(2\pi\hbar^{2}/mk_{B}T\right)^{1/2}$ is the boson thermal-wavelength.  
There is another characteristic temperature for which the boson thermal wavelength is less than or equal to approx. 0.7 $a_{x}$ that marks the point at which the effects of the walls are less conspicuous, above this temperature the system shows the usual 3D free IBG behavior. The onset of this 3D behavior is observed as the last maximum on the right in the specific heat curve. 

As $P_{0}$ becomes larger, the specific heat of the isotropic case reveals a \emph{one-dimensional} behavior in a particular range of temperatures determined by the length scale $a,$ as is shown in Fig. \ref{fig:CvP05000}. In both cases, $a_{x}/\protect\lambda _{0} = a_{y}/\protect%
\lambda _{0} = 0.5,$ and $0.1$ the specific heat $C_{V}/Nk_{B},$ approaches the one-dimensional classical value $1/2$ over a relatively large region of temperatures. This behavior is more pronounced as $P_{0}$ is
increased. 

In Fig. \ref{fig:Cv2} we can see that for the {\it anisotropic} case where $a_x\neq a_y$, the BEC transition occurs at a temperature between the respective critical temperatures for the isotropic cases $a=\hbox{max}\{a_{x},a_{y}\}$ and $a=\hbox{min}\{a_{x},a_{y}\}$, and that the minimum corresponding to each isotropic case, both appear in the anisotropic one.
%
\begin{figure}[tbh]
\centerline{%
\epsfig{file=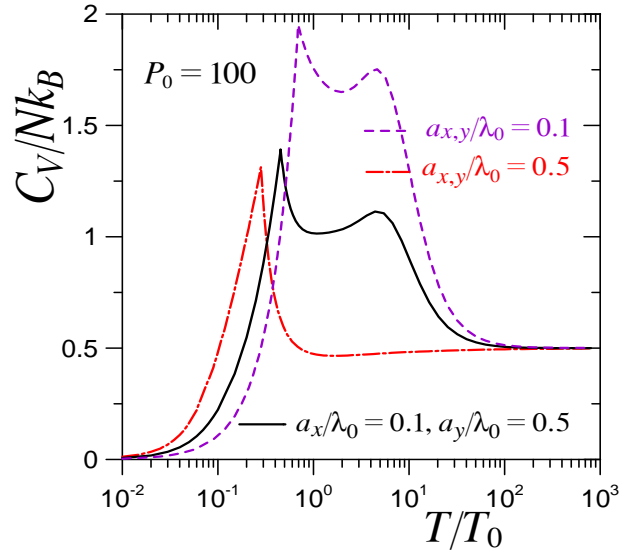,height=2.9in,width=3.3in}}
\caption{(Color online) Specific heat in $Nk_{B}$ units, as a function of $%
T/T_{0}$, for $a_{x}/\protect\lambda _{0}=0.5$ and $a_{y}/\protect\lambda %
_{0}=0.1$ and $P_{0}=100$ (black line) compared to their respective isotropic cases: $%
a_{x}/\protect\lambda _{0}=a_{y}/\protect\lambda _{0}=0.1$ and $0.5$, (color lines). One band only.}
\label{fig:Cv3}
\end{figure}

\subsection{Density of states}
The effects of the band structure are conspicuously exhibited in the density of states (DOS)
\begin{equation}
g(\epsilon)=\sum_{k_{x},k_{y},k_{z}}\delta(\epsilon-%
\varepsilon_{k_{x}}-\varepsilon_{k_{y}}-\varepsilon_{k_{z}}),
\end{equation}
which can be written in the thermodynamic limit as
\begin{multline}
g(\epsilon)=\frac{L^{3}}{(2\pi)^{3}} \sum_{j_{x},j_{y}=1}^{\infty}\int_{-%
\pi/a_{x}}^{\pi/a_{x}}dk_{x}\int_{-\pi/a_{y}}^{\pi/a_{y}}dk_{y} \\
\int_{-\infty}^{\infty}dk_{z}\,
\delta(\epsilon-\varepsilon_{k_{x},j_{x}}-\varepsilon_{k_{y},j_{y}}-%
\varepsilon_{k_{z}}),
\end{multline}
where we have explicitly written the integrals over the energy-spectrum in the $x$ and $y$ directions as a sum over bands of the integrals over $k_{x}$ and $k_{y}$ in the first Brillouin zone.

Upon integration over $dk_{z}$ we obtain
\begin{multline}  \label{DOS}
g(\epsilon)=\frac{L^{3}}{(2\pi)^{3}}\left(\frac{2m}{\hbar^{2}}%
\right)^{1/2} \sum_{j_{x},j_{y}=1}^{\infty}
\int_{-\pi/a_{x}}^{\pi/a_{x}}dk_{x} \\
\int_{-\pi/a_{y}}^{\pi/a_{y}}dk_{y} \frac{\theta(\epsilon-%
\varepsilon_{k_{x}j_{x}}-\varepsilon_{k_{y}j_{y}})}{\sqrt{%
(\epsilon-\varepsilon_{k_{x},j_{x}}-\varepsilon_{k_{y},j_{y}})}},
\end{multline}
where $\theta(x)$ is the Heaviside step function. For energies close to the
minimum, $g(\epsilon)$ varies as $\epsilon^{1/2}$ as it does the DOS
of a free particle in three dimensions. This can be shown by noting that for
energies close to $\varepsilon_{0i}$ and for small $k$, expression (\ref{eqWK}%
) can be approximated by $\varepsilon_{i0} + {\hbar^{2}k_{i}^{2}}/2M_{i}$.
In the isotropic case ($\varepsilon_{x0}= \varepsilon_{y0}= \varepsilon_{0}/2$ and $M_{x} = M_{y} = M$ ), we can therefore write
\begin{multline*}  \label{DOS2}
g(\tilde{\varepsilon}) \simeq \frac{L^{3}}{(2\pi)^{3}}\left(\frac{2m}{%
\hbar^{2}}\right)^{1/2} 2\pi\int_{0}^{\tilde{k}}dk\, k \frac{\theta(\tilde{%
\varepsilon}-\frac{\hbar^{2}}{2M}k^{2})}{\sqrt{\tilde{\varepsilon}-\frac{%
\hbar^{2}}{2M}k^{2}}},
\end{multline*}
where $\tilde{\varepsilon}\equiv\epsilon - \varepsilon_{0},$ $%
k^{2}=k_{x}^{2}+k_{y}^{2}$ and $\tilde{k}$ is the maximum value of $k$ in
the first Brillouin zone. The exact value is not needed as long as $\tilde{%
\varepsilon}<\left(\hbar^{2}/2Ma^{2}\right)\tilde{k}^{2}.$ Thus, after
evaluating the integral by a change of variable we obtain
\begin{equation}  \label{DOS3}
g(\tilde{\varepsilon}) \simeq \frac{L^{3}}{(2\pi)^{2}}\left(\frac{2m}{%
\hbar^{2}}\right)^{1/2}\frac{2M}{\hbar^{2}}\tilde{\varepsilon}^{1/2}.
\end{equation}
\begin{figure}[tbh]
\epsfig{file=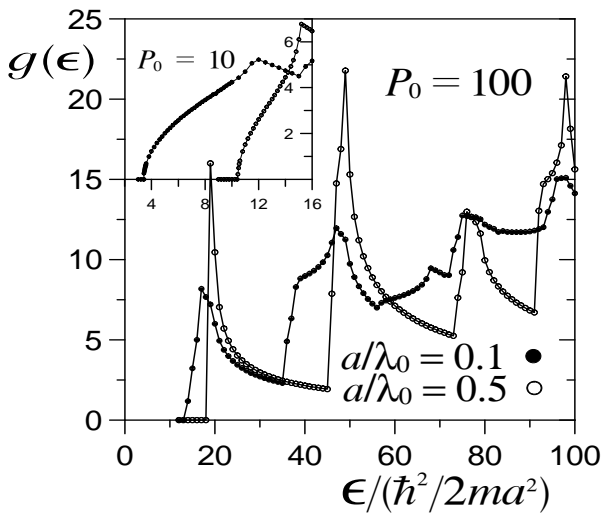,height=2.7in,width=3.2in}
\caption{Density of states for tubes with $P_0 = 100$, $a/\protect\lambda_0
= 0.1$ (full dots) and $a/\protect\lambda_0 = 0.5$ (open dots).}
\label{fig:g(e)}
\end{figure}
The DOS $g(\epsilon)$ from Eq. (\ref{DOS}) is plotted in Fig. \ref{fig:g(e)} for $P_0 = 100$, $%
a_x/\lambda_0 = a_y/\lambda_0 = 0.1$ and 0.5. For energies around the bottom of the first band
given by (\ref{DOS3}) is shown in the inset where a smaller value of $P_{0}$ has been used to reveal
the 3D behavior. The $P_0$ dependence of $g(\epsilon)$ enters only
through the effective mass $M,$ and in general, $g(\epsilon)$ is a rather
complex function of the energy (full-doted curve), however, in the limit of
large values of $P_{0}$ the reminiscent behavior $\epsilon^{-1/2}$ of the
one-dimensional DOS of a free particle is observed as a
``falling'' (see for instance the curve in open-circles) of $g(\epsilon)$
with $\epsilon.$ This behavior occurs when the energy-states along the $z$
direction are the only ones that contribute to $g(\epsilon),$ the
contribution due to the other two directions being a constant related to the
energy band-gaps.

\section{Discussion and conclusions}

At this point let us make the following remarks. In Fig. \ref{fig:Cv3} we plot $C_V$ as a function of $T/T_0$ for different plane separations and $P_0=100$, where \emph{only} the first energy band, exactly computed from Eq. (\ref{KPRD}) in the $x$- and $y$-directions, has been considered. This situation qualitatively resembles the results obtained if the dispersion relation (\ref{eqWK}) were used\cite{haerdig,wenkan}.  We consider the corresponding calculation of $C_V$ in Fig. \ref{fig:Cv2}, where we have used enough number of exact energy bands needed to achieve the required precision in the calculation of this property. We note that both curves coincide quantitatively and qualitatively for temperatures not greater than $T_0$; for temperatures between $T_0$ and about 5 $T_0$ the qualitatively behavior remains best for the smaller plane separations; finally, for temperatures greater than 5 $T_0$ even the qualitative agreement disappears. Important structural information is missing when just one band is included in the calculation, for instance, the second minimum that appears in the isotropic case of Fig. \ref{fig:Cv2} (full-black line) is missing in the corresponding curve in Fig. \ref{fig:Cv3}. Clearly, the use of the lowest band leads to a one-dimensional behavior in the classical limit even though the system is three-dimensional.

In the isotropic case (that can be related to homogeneous nanotubes bundle-system),
our model predicts a one-dimensional behavior of the specific heat in a
range of temperatures determined by the distance between the delta-barriers.
Such results shown in Fig. \ref{fig:CvP05000} can be used to comparatively
explain the one dimensional character of the specific heat of adsorbed $^{4}$%
He in single-wall nanotubes as expected and observed in experimental
situations\cite{Lasjaunias2003}. Although Lasjaunias et al. did not
report the presence of the peak that marks the BEC phase-transition we
explain this fact in terms of the finite size of the experimental system
where such transition can be made much less conspicuous.

One important result is that our system always exhibits a BEC which is
possible due to the coupling between the different channels the bosons move
in. When the wall impenetrability goes to infinity our system becomes a set of decoupled tubes with zero BEC critical temperature as expected. Some results,  
often found in the literature, which claim that
BEC in a bundle of homogeneous nanotubes is not possible\cite{Gatica}, agree with ours if homogeneous means non-communication among tubes. However in the
experimental set up, it is clear that the heterogeneity of the interstitial
channels may have strong effects on the thermodynamic behavior of the
system, in particular in the BEC critical temperature which can be different from zero if the heterogeneity of the channels causes exchange of atoms among them. 


On the other hand, the use of multifilamentary superconducting tapes \cite{Sumiyoshi} or wires\cite{Junod2007} has improved the coil performance to support higher critical current density useful to create higher magnetic field to be used, for example, in the Large Hadron Collider currently under operation or the
International Thermonuclear Experimental Fusion Reactor planed to work in 2019. In both cases tapes and wires
gather many filaments where pairs flow preferentially along the longitudinal
direction. Critical temperature distribution in the Nb$_3$Sn strands as well
as the specific heat have been reported. The authors point out two
transitions in the specific heat curve: one at the critical temperature
(around 18 K) they associate to the complex superconductor wire and another
one at a lower temperature (around 9 K) they associate to the unreacted Nb.
However, if the Cooper pairs should be considered as bosons, our specific
heat calculations show at least two critical temperatures: the lower one
associated to the collective effect of the filaments (our tubes) and a
smooth maximum corresponding to the boson gas behavior in the individual
filaments. In other words, the meaning given to the maxima in our Fig. \ref%
{fig:Cv2} and in the Fig. 4 of Ref. [\onlinecite{Junod2007}], are
interchanged. A way to elucidate this controversy would be to make wires
with larger filament diameters, then prove that their smooth maximum shifts
to the left as it is observed in our calculations.



We summarize our main results in the following list:
we observe that in the presence of periodical structures constructed with orthogonal Dirac combs inside an infinite box filled with bosons, the
critical temperature decreases from the 3D ideal boson gas $T_0$ as $P_0$ increases, while the plane separations $a_{x,y}/\lambda _{0}$ are kept constant. It becomes zero when the periodic delta potential strengths become
infinite. In other words, there is not BEC critical temperature different from zero for bosons in a tube of finite cross section and infinite length.

As the separation between planes is lowered, the critical temperature reaches a $P_0$-dependent minimum value and then it is expected \cite{PatyJLTP,PatyPRA} to increase again towards $T_0$.

For systems with $a_{x,y} > \lambda_0$ the numerical calculations for the
critical temperature and specific heat are very sensitive to the number of
energy bands considered. To attain convergence we need to include up to 1000
bands.

At $T=T_c$, the specific heat is continuous but has a discontinuity in its
derivative. In the isotropic case it has one minimum and one or two maxima.
The minimum is associated to particle trapping between two planes when its
thermal wavelength is equal to $2a$. This is corroborated in the anisotropic
case where the specific heat shows not one but \textit{two} minima
associated with the particle trapping in the $x$ or $y$ directions.

The maximum at higher temperatures is associated to the onset of the
system's approach to a 3D IBG behavior in this regime where the thermal
wavelength $\lambda \lesssim 0.7a$.

While there is still a controversy over whether or not Bose-Einstein condensation of $^{4}$He exists inside interstitial filaments in bundles of carbon nanotubes, we conclude that in order to have BEC there must be a way through which the interstitial channels are coupled among them, either by effects of inhomogeneity of the tube bundles or by tunneling across the weaker interstitial walls. 

Finally, we mention that the proposed model in this paper gives account of systems composed of a very
large number of quasi-one dimensional systems such as: bundles of carbon
nanotubes, superconductor-multistrand wires, Bechgaard salt or 2D
opto-magnetic traps.\\

We acknowledge the partial support from grant PAPIIT IN105011, IN117010-3,
and IN106908. CONACyT 104917.

\end{document}